\documentclass[prd,aps,showpacs,
nofootinbib,%
tightenlines,
notitlepage,
twocolumn
]{revtex4-1}

\usepackage{graphicx}
\usepackage{amsmath}
\usepackage{amssymb}
\usepackage{hyperref} 
\usepackage{upgreek}

\newcommand{\be}{\begin{equation}}
\newcommand{\ee}{\end{equation}}
\newcommand{\bea}{\begin{eqnarray}}
\newcommand{\eea}{\end{eqnarray}}
\newcommand{\nn}{\nonumber}

\begin{document}

\title{Hadronic light-by-light contribution to 
the muon $g-2$\\
from holographic QCD with solved $U(1)_A$ problem}

\author{Josef Leutgeb}
\author{Jonas Mager}
\author{Anton Rebhan}
\affiliation{Institut f\"ur Theoretische Physik, Technische Universit\"at Wien,
        Wiedner Hauptstrasse 8-10, A-1040 Vienna, Austria}

\date{\today}

\begin{abstract}
We employ the comparatively minimal extension of hard-wall AdS/QCD due to Katz and Schwartz which takes into account the U(1)$_A$ anomaly for computing hadronic light-by-light scattering contributions of pseudoscalar and axial vector mesons to the anomalous magnetic moment of the muon $a_\upmu$. By including a gluon condensate as one extra tunable parameter besides those fixed by $f_\pi$ and the pion, kaon, and rho masses, we obtain remarkably accurate fits for $\eta$ and $\eta'$ masses and their decay rates to photons, leading to $a_\upmu$ contributions in complete agreement with the Standard Model result by the Muon $g-2$ Theory Initiative. Turning to the less well understood axial vector contributions, we update our previous predictions obtained in flavor-symmetric hard-wall AdS/QCD models without U(1)$_A$ breaking.
\end{abstract}

\maketitle

\section{Introduction}

Since in 2021 the Muon $g-2$ Collaboration at Fermilab
\cite{Muong-2:2021ojo}
has succeeded in confirming and improving the result
of the E821/BNL measurement from 2006 \cite{Muong-2:2006rrc}
for the anomalous magnetic moment of the
muon \cite{Jegerlehner:2017gek}
and is under way on further increasing its accuracy,
the existing uncertainties in the disagreeing theoretical Standard Model
result \cite{Aoyama:2020ynm} need to be scrutinized and
also improved.

Whereas QED \cite{Aoyama:2012wk,*Aoyama:2017uqe,*Aoyama:2019ryr} and electroweak contributions \cite{Czarnecki:2002nt,Gnendiger:2013pva} are
sufficiently under control, the theoretical uncertainty
is dominated by hadronic effects
\cite{Melnikov:2003xd,Prades:2009tw,Kurz:2014wya,Colangelo:2014qya,Pauk:2014rta,Davier:2017zfy,Masjuan:2017tvw,Colangelo:2017fiz,Keshavarzi:2018mgv,Colangelo:2018mtw,Hoferichter:2019gzf,Davier:2019can,Keshavarzi:2019abf,Hoferichter:2018dmo,*Hoferichter:2018kwz,Gerardin:2019vio,Bijnens:2019ghy,Bijnens:2020xnl,Colangelo:2019lpu,Colangelo:2019uex,Danilkin:2019mhd,Blum:2019ugy,Chao:2021tvp,Hoferichter:2021wyj,Danilkin:2021icn}.
The largest contribution by far is the hadronic vacuum polarization (HVP),
where a recent lattice calculation \cite{Borsanyi:2020mff}
is at variance with the result of the 2020 White Paper (WP) of the Muon $g-2$ Theory Initiative \cite{Aoyama:2020ynm} beyond the
respective estimated errors, leading to a less strong deviation
from the experimental result if the lattice result is used
in place of the data-driven one obtained in the WP.
Once this discrepancy is resolved, it will be important to
also reduce the uncertainty in the contribution from hadronic light-by-light scattering (HLBL), which at present has errors at the level of 20\%,
which in absolute numbers are comparable to the small errors aimed for in the case
of HVP.

Besides the dominant pion pole contribution to HLBL, which by now
seems to be well understood, and where data-driven approaches and
lattice evaluations agree perfectly, and the similarly well determined contributions from $\eta$ and $\eta'$ mesons, other single-meson contributions
are much less under control. An important contribution is
expected in particular from axial vector mesons, which like
pseudoscalars have anomalous couplings to photons. However,
theoretical predictions from various hadronic models vary
a lot \cite{Bijnens:2001cq,Melnikov:2003xd,Pauk:2014rta,Jegerlehner:2017gek,Roig:2019reh,Masjuan:2020jsf}, which has led to a WP estimate
of the axial vector contribution with 100\% uncertainty.

Holographic QCD models motivated by the AdS/CFT correspondence \cite{Maldacena:1997re,Witten:1998zw,Aharony:1999ti}
have proved to be remarkably successful in qualitatively and also
quantitatively describing hadronic observables,
even those with a minimal set of
parameters and the simplest geometry of anti-de Sitter space with a hard-wall (HW) cutoff.
Such AdS/QCD models are not good enough to help with the
current discrepancy between different predictions for the HVP contribution, where sub-percent accuracy is required. However,
they are certainly of interest for estimating the HLBL contributions.

In Ref.~\cite{Leutgeb:2019zpq}, we have revisited previous
studies \cite{Hong:2009zw,Cappiello:2010uy} of the pion pole contribution to HLBL and its consequences
for the value of $a_\upmu=(g-2)_\upmu/2$ using simple bottom-up AdS/QCD models
in the chiral limit and we have found a satisfactory agreement
with the data-driven and lattice approaches. The transition
form factors obtained in AdS/QCD involve infinite towers of
vector mesons, realizing vector meson dominance (VMD) in a form
that is consistent with the asymptotic behavior derived
from perturbative QCD \cite{Hoferichter:2020lap}
for both, the singly and the doubly virtual case.

In \cite{Leutgeb:2019gbz,Cappiello:2019hwh}, also the
contribution from the infinite tower of axial vector mesons and
their anomalous coupling to photons has been calculated,
and it could be shown that this takes care of the
long-standing problem that simpler hadronic models had
with the Melnikov-Vainshtein (MV) constraint \cite{Melnikov:2003xd}
on the HLBL scattering
amplitude (see \cite{Colangelo:2021nkr} for
a review assessing its impact on $a_\upmu$).
In \cite{Leutgeb:2021mpu}, we have more recently
extended these calculations to include finite quark masses
in the flavor-symmetric case. Besides demonstrating that
the saturation of the MV constraint is entirely due to
axial vector mesons also away from the chiral limit, we have
confirmed the relatively large contribution obtained in
the chiral model.

In the present paper, we consider a minimal extension of the original
hard-wall AdS/QCD model \cite{Erlich:2005qh} due to Katz and Schwartz \cite{Katz:2007tf} for solving the U(1)$_A$ problem associated with 
the relatively large $\eta'$ mass. 
Going slightly beyond the setup of \cite{Katz:2007tf} by
including a nonvanishing gluon condensate, we find that a very accurate
match of the masses of $\eta$ and $\eta'$ mesons as well as their
coupling strength to photons can be achieved.
We then use this model to evaluate all contributions of
pseudoscalar\footnote{The pseudoscalar contributions to $a_\upmu$ have been
calculated before in \cite{Hong:2009zw} by Hong and Kim in the
Katz-Schwartz model without gluon condensate. As discussed below,
we disagree in the treatment of the Chern-Simons term.} and 
axial vector meson excitations, and thereby also the effect of the MV short-distance constraint, to the HLBL contribution to $a_\upmu$.

\section{Katz-Schwartz model: Hard-wall AdS/QCD with solved $U(1)_A$ problem}

The model proposed by Katz and Schwartz \cite{Katz:2007tf,Schafer:2007qy} for solving the $U(1)_A$ problem builds upon the original HW AdS/QCD models of Ref.~\cite{Erlich:2005qh,DaRold:2005mxj} which have turned out to
provide a remarkably good approximation to the physics of light hadrons
while introducing a minimal set of parameters.

In these models, one keeps the background geometry of pure anti-de Sitter space
with metric
\be
ds^2=z^{-2}(\eta_{\mu\nu}dx^\mu dx^\nu - dz^2),
\ee
cut off by a hard wall at a finite value of the holographic radial coordinate at $z=z_0$ with suitable boundary conditions for the five-dimensional fields
that at the conformal boundary at $z=0$ represent sources for a set of
QCD operators of interest.
In addition to five-dimensional Yang-Mills fields $\mathcal{B}^{L,R}$ dual to left and right chiral
quark currents, a bifundamental scalar $X$ representing quark-antiquark bilinears is introduced for spontaneous symmetry breaking of $U(N_f)\times U(N_f)\to U(N_f)_V$.
Confinement is implemented by
the cutoff at $z_0$,
where boundary conditions for the five-dimensional fields are imposed.

The five-dimensional Yang-Mills action
\bea
S_{\rm YM} &=& -\frac{1}{4g_5^2} \int d^4x \int_0^{z_0} dz\,
\sqrt{g}\, g^{PR}g^{QS}
\nonumber\\
&&\qquad\text{tr}\left(\mathcal{F}^\mathrm{L}_{PQ}\mathcal{F}^\mathrm{L}_{RS}
+\mathcal{F}^\mathrm{R}_{PQ}\mathcal{F}^\mathrm{R}_{RS}\right),
\eea
where $P,Q,R,S=0,\dots,3,z$ and $\mathcal{F}_{MN}=\partial_M \mathcal{B}_N-\partial_N \mathcal{B}_M-i[\mathcal{B}_M,\mathcal{B}_N]$, is
augmented by
a Chern-Simons action $S_{\rm CS}=S_{\rm CS}^\mathrm{L}-S_{\rm CS}^\mathrm{R}$ 
to account for flavor anomalies, reading (in
differential form notation)
\be\label{SCS}
S_{\rm CS}^\mathrm{L,R}=\frac{N_c}{24\pi^2}\int\text{tr}\left(\mathcal{B}\mathcal{F}^2-\frac{i}2 \mathcal{B}^3\mathcal{F}
-\frac1{10}\mathcal{B}^5\right)^\mathrm{L,R},
\ee
(up to a boundary term at $z_0$ that needs to be subtracted \cite{Grigoryan:2007wn,Leutgeb:2021mpu}).

The bifundamental bulk scalar $X$ is
parametrized as
\cite{Abidin:2009aj}
\be
X=e^{i\eta^a(x,z) t^a}X_0e^{i\eta^a(x,z)t^a},
\ee
where $\eta^a$, $a=0,\ldots,8$, is a nonet of pseudoscalars excitations.
The five-dimensional mass of $X$ is fixed at\footnote{In \cite{Leutgeb:2021mpu}
we have also studied the generalization to other values of $M_X$ as
proposed in \cite{Domenech:2010aq}.}
$M_X=-3$ by the scaling
dimension of the dual operator $\bar q_L q_R$,
leading to a vacuum solution
\be
X_{0ij}=\frac12 m_{ij}z+\frac12 \sigma_{ij}z^3.
\ee
Choosing $N_f=3$, we restrict ourselves to the isospin symmetric case
$m_u=m_d=m_q\not=m_s$ with
$X=\frac12\text{diag}(v_q,v_q,v_s)$.

For taking care of the $U(1)_A$ problem, a massless complex field $Y$ is
introduced, representing the gluon field strength squared $\alpha_s G_{\mu\nu}^2$ by its modulus and $\alpha_s G\tilde G$ by its phase, such that the Lagrangian of scalars reads
\bea
&&\mathcal{L}_{X,Y}/\sqrt{g}=
\text{tr}\left[ |DX|^2+3|X|^2\right]\nonumber\\
&&+\frac1{2(\ln z\Lambda)^2}|DY|^2+\frac{\kappa_0}{4}\left[\bar{Y}^{N_f}
\det(X)+\text{h.c.}\right],\label{LXY}
\eea
where  the logarithm in front of the kinetic term for $Y$ accounts for
the fact that its dual operators approach scaling dimension 4 only
asymptotically.
The complex scalar field $Y$ is charged only under the singlet axial vector field and hence its coupling is given by
\bea
 D_M Y= \partial_M Y+\frac{i}{\sqrt{2 N_f}}(\mathcal{B}^{L,0}_M-\mathcal{B}^{R,0}_M)Y.
\eea
Without the logarithm in \eqref{LXY}, the field equations for $Y$
would give a background $\langle Y\rangle = C + \Xi z^4$, where
$\Xi$ represents a gluon condensate.
After absorbing some numerical factors into $C$, the authors of \cite{Katz:2007tf} use the axial anomaly relation and the QCD operator product expansion (OPE fit) of the flavor-singlet axial vector correlation function to find
\be
C=\frac{\alpha_s}{2\pi^2}\sqrt{2N_f}.
\ee
In QCD $\alpha_s$ is a running coupling and since in holography energy is identified with $z^{-1}$, they argue that $\alpha_s$ should be made $z$-dependent.

Matching $z\partial_z\alpha_s(z)\simeq -\beta_0\alpha_s^2$ with the one-loop
QCD $\beta$ function where $\beta_0=9/2\pi$ for $N_c=N_f=3$
gives $\alpha_s^{-1}=\beta_0\ln(\Lambda z)$, which
is adopted for all $z<z_0=\Lambda^{-1}$. 
Making $\alpha_s$ and therefore $C$ depend on $z$ is of course inconsistent with the equations of motion (without the logarithm), hence we use the modified version \eqref{LXY} including the logarithms.
The presence of the logarithm in the action modifies the field equations for $Y$, leading to the general solution%
\footnote{Here we also deviate from Ref.~\cite{Katz:2007tf},
where a gluon condensate, here parametrized by $D_1$ was neglected and only $C$ of the background solution had to be modified.}

\be\label{yvev}
\langle Y\rangle=D_0 +D_1 z^4 \left[(\ln z\Lambda)^2 -\frac12\ln z\Lambda+\frac{1}{8} \right].
\ee
For later convenience we define\footnote{This is the same redefinition that \cite{Katz:2007tf} use, except for the logarithm} $\tilde{Y}_0=\frac{2}{\sqrt{2 N_f}}(-\ln z \Lambda)^{-1}\langle Y\rangle 
$, which is parametrized as
\be
\tilde{Y}_0=\frac{C_0}{-\ln z \Lambda}-\Xi_0 z^4 \big( (\ln z\Lambda) -\frac{1  }{2}+\frac{1}{8\ln z \Lambda } \big)
\ee
with $C_0=\sqrt{2N_f}/(2\pi^2\beta_0)=\sqrt{2/3}/(3\pi)$.
This background now naturally incorporates the running of $\alpha_s$ consistently and permits a nonvanishing gluon condensate through nonzero values of $\Xi_0$.

The coupling constant $g_5$ can be fixed by the OPE of the
vector current correlator as
\be\label{g5LO}
g_5^2=12\pi^2/N_c=(2\pi)^2\quad \text{(OPE fit)},
\ee
but we shall alternatively consider matching the
decay constant of the $\rho$ meson,
which in the hard-wall model leads to \cite{Leutgeb:2022cvg}
\be\label{g5Frho}
g_5^2=0.894(2\pi^2)\quad \text{($F_\rho$-fit)}.
\ee
The latter leads to a significant improvement of the holographic
result for the hadronic vacuum polarization:
With the leading-order OPE fit \eqref{g5LO}, there is a deviation
of 14\% from $N_f=2$ dispersive results, which is reduced to
about 5\% with \eqref{g5Frho} \cite{Leutgeb:2022cvg}.
A reduction of $g_5^2$ by about 10\% appears to be
warranted also by comparing with next-to-leading order QCD results
for the vector correlator at moderately large $Q^2$ values \cite{Shifman:1978bx,Melic:2002ij,Bijnens:2021jqo}.
It also brings our $N_f=2$ results for the pion pole contribution to $a_\upmu$
\cite{Leutgeb:2021mpu} in line with the WP result.

With $g_5$ and $C_0$ fixed by the UV behavior, the free parameters of the model
are (i) the location of the hard wall, $z_0$, which can be identified with $\Lambda^{-1}$, and will be set by the $\rho$ meson mass,
(ii) quark masses in $m_{ij}=\text{diag}(m_q,m_q,m_s)$,
(iii) chiral condensates $\sigma_{ij}$, which we shall assume to be
given by a single parameter, $\sigma_{ij}=\sigma\delta_{ij}$,
and (iv) $\Xi_0$, which corresponds to the gluon condensate $\alpha_s\langle G^2\rangle$.
The coupling constant $\kappa_0$, on the other hand, can be
set to some sufficiently large value, since it turns out that
for $\kappa_0\gg1$ all results depend only weakly on $\kappa_0$ \cite{Katz:2007tf}.

\section{Meson modes and transition form factors}

Vector meson dominance is naturally part of this model by
relating a nonzero boundary value $\mathcal{B}^V_\mu(0)=e\mathcal{Q}A_\mu^\mathrm{e.m.}$ 
to the background electromagnetic potential and setting
$\mathcal{Q}=\text{diag}(\frac23,-\frac13,-\frac13)$ according to the charges of up, down, and strange quarks.
Normalizable modes of $\mathcal{B}_V$ and $\mathcal{B}_A$
correspond to vector and axial vector mesons, the longitudinal polarizations
of the latter mixing with the pseudoscalars $\eta^a$ in $X$.
Ignoring scalar excitations, which in this model do not
couple to photons,\footnote{See Ref.~\cite{Cappiello:2021vzi} for extensions of the HW model where further interactions are switched on to have scalars couple to photons in order to study their potential contribution to $a_\upmu$ in AdS/QCD.} the additional $Y$ field also involves a
pseudoscalar $a$ through its phase
\be
Y=\langle Y\rangle \exp\left[{i2a(x,z)/\sqrt{2N_f}}\right],
\ee
which corresponds to a pseudoscalar glueball in the boundary theory ($G$) that couples to photons via its mixing with the flavor-singlet pseudoscalar mesons.

To determine the pseudoscalar eigenmodes in the mixed $a=(0,8)$-sector, we consider the equations of motion\footnote{We assume a summation over $a=(0,8)$ for contracted flavor indices and work in the $A_z=0$ gauge.}
\bea
\label{eq:phi}
     &&\partial_z\left(\frac{1}{z} \partial_z \varphi^a_n \right) + g_5^2 \frac{M_{ab}^2}{z^3} \left(\eta^b_n-\varphi^b_n\right) \nn\\
     &&\qquad+\delta^{a0}  g_5^2 \frac{\tilde{Y}_0^2}{z^3} \left(a_n-\varphi^0_n\right)=0,
\eea
\bea
\label{eq:aeq}
    &&\partial_z\left(\frac{\tilde{Y}_0^2}{z^3} \partial_z a_n \right)  + m_n^2 \frac{\tilde{Y}_0^2}{z^3} \left(a_n-\varphi^0_n\right)   \nn\\
    &&\qquad+ { \kappa N_f}  \frac{v_q^2 v_s }{z^5 } \left(\frac{ \tilde{Y}_{00}}{4}\right)^{N_f}  \left(\eta^0_n-a_n\right)=0,
\eea
\bea
    &&\frac{m_n^2}{g_5^2}  \frac{1}{z} \partial_z \varphi^a_n-  \delta^{a0}\frac{{\tilde{Y}_0^2}}{z^3} \partial_z a_n     -\frac{M^2_{ab}}{z^3}\partial_z\eta^b_n=0,
\eea
with the longitudinal component of the axial gauge field $\partial_\mu \varphi^a=A^{a\parallel}_\mu$ and an effective 5-dimensional mass term
\begin{equation}
    M^2_{ab} = \frac{1}{3}
    \begin{pmatrix}
        2 v_q^2+ v_s^2 & \sqrt{2}( v_q^2-v_s^2)\\
         \sqrt{2}(v_q^2-v_s^2) & v_q^2+ 2 v_s^2
    \end{pmatrix},
\end{equation}
with $v_{q,s}=m_{q,s} z + \sigma z^3$.
Here we have also absorbed numerical constants in $\kappa_0$ and renamed it to $\kappa$, and we defined $\tilde{Y}_{00}=-{\tilde{Y}_0 \ln z \Lambda}/{C_0}$.

The fields $\varphi^a$ are dual to QCD operators $-\partial_{\mu}J_A^{\mu, a}$, and the glueball field $a$ is dual to $-\sqrt{2N_f}K$, where K is the instanton density $K=\frac{\alpha_s}{8 \pi}G^a_{\mu \nu}\tilde{G}^{a\mu \nu}$. This new field $a$ allows then among other things to compute overlaps of the instanton density $K$ with pseudoscalar modes $\eta, \eta',...$ and the topological susceptibility.

All fields of the normalizable modes have Dirichlet boundary conditions in the UV at $z=0$, Neumann boundary conditions in the IR at $z=z_0$, and are canonically normalized by
\begin{equation}
    \int dz \left( \frac{M_{ab}}{z^3}(\eta^a_n(\eta^b_m-\varphi^b_m))+\frac{\tilde{Y_0}^2}{z^3}a_n(a_m-\varphi^0_m) \right) = \delta_{nm}.
\end{equation}

From the Chern-Simons term \eqref{SCS} we obtain the transition form factor (TFF)
\bea
\label{eq:TFFdecomp}
    &&F_n(Q_1^2, Q_2^2)=\text{tr}(t^a \mathcal{Q}^2) F_n^{a}(Q_1^2, Q_2^2),
\eea
with \cite{Leutgeb:2021mpu}
\bea\label{CSsubtraction}
    &&F_n^a(Q_1^2, Q_2^2)= -\frac{N_c}{2 \pi^2} \bigg( \int dz \varphi'^a_n(z) \mathcal{J}(Q_1,z) \mathcal{J}(Q_2,z) \nn\\
    &&\qquad-\left[\varphi^a_n(z)-\eta^a_n(z)\right] \mathcal{J}(Q_1,z) \mathcal{J}(Q_2,z)\Big|_{z_0} \bigg),
\eea
where the vector bulk-to-boundary propagator
\be\label{HWVF}
    \mathcal{J}(Q,z)=
    Qz \left[ K_1(Qz)+\frac{K_0(Q z_0)}{I_0(Q z_0)}I_1(Q z) \right]
\ee
describes virtual photons 
with spacelike momentum $q^2=-Q^2$.

Note that (\ref{CSsubtraction}) involves the subtraction
of a boundary term at $z=z_0$, which is absent in \cite{Katz:2007tf}, but (\ref{CSsubtraction}) also differs from the
corresponding expression given in \cite{Grigoryan:2007wn},
where $(\varphi^a_n-\eta^a_n)'$ appears in the integral.
As discussed in \cite{Leutgeb:2021mpu}, this is only correct
in the chiral limit and for the ground-state pion, but not
in the massive case. Our results for the pseudoscalar TFFs therefore also differ from Ref.~\cite{Hong:2009zw}, where
the TFFs for ground-state pseudoscalars in the Katz-Schwartz model
have been evaluated with the inapplicable chiral formula of \cite{Grigoryan:2007wn}.

We can also generalize the asymptotic results and the
sum relations obtained in \cite{Leutgeb:2021mpu} 
in the nonchiral but flavor-symmetric case
to the asymmetric case $m_s\not=m_q$ with broken U(1)$_A$ symmetry. Most importantly, we can derive the anomaly equations
\begin{equation}
\label{eq:anomeq}
    \sum_n f_n^a F_n^{a}(0,0) = \frac{N_c}{2 \pi^2}, \quad a=0,3, \,\text{or}\; 8 \;\text{(fixed)},
\end{equation}
and the short-distance constraint (SDC)
\begin{equation}
\label{eq:SDC}
    F_n^a\left(Q^2 (1+w), Q^2 (1-w)\right)\to \frac{N_c}{2 \pi^2} g_5^2 f_n^a \frac{1}{Q^2} f(w)
\end{equation}
 for $Q \to \infty$ with the asymmetry function
\begin{equation}
    f(w)=\frac{1}{w^2}-\frac{1-w^2}{2w^3}\ln\frac{1+w}{1-w},
\end{equation}
and the pseudoscalar decay constants 
\be
f_n^a=\left.-g_5^{-2}\partial_z\varphi^{a}_n/z\right|_{z\to0}. 
\ee
Note that the decay rate associated with the glueball field, 
\be
\left. f_G^n=\tilde{Y}_0^2 \partial_z a_n / z^3\right|_{z\to0}
\ee
does not contribute to the TFF. In QCD $ f_G^n$ computes $(-\sqrt{2 N_f})\langle \Omega|K|P_n\rangle$, where $P_n$ is the respective pseudoscalar particle.

As far as we know, the sum rules \eqref{eq:anomeq} have not appeared in the
literature before in this general form where they include mixing due to finite
quark masses $m_s\not= m_q$ and breaking of the U(1)$_A$ symmetry at finite $N_c$.
They involve the components $F_n^a$ of the TFFs $F_n$ defined by \eqref{eq:TFFdecomp},
which at least in lattice QCD could be determined directly 
by varying the quark charge matrix
$\mathcal{Q}$. 

In general, all modes contribute to \eqref{eq:anomeq}. In the
limit of vanishing quark masses, one has $m_n^2 f_n^{3,8}=O(m_q)$ for each $n$ (as
discussed in the appendix of \cite{Leutgeb:2021mpu} in the holographic setup).
This implies that for $a=3$ and 8 only the massless Goldstone bosons contribute to the sum rules, whereas excited pseudoscalar modes decouple from the anomaly relations, while they can still have nonzero $F_n(0,0)$. In the $a=0$ sector, we instead find $f^0_n m_n^2+f_{G}^n=O(m_q)$,
and the mass of the $\eta'$ meson is nonzero (in the chiral limit between $600$ and $700$ MeV, depending on the precise value of $\Xi_0$). This means that even in the chiral limit \eqref{eq:anomeq} receives contributions from all higher modes in the $a=0$ sector, which is qualitatively different from the $a=3,8$ sectors.

Comparing the axial vector sector to \cite{Leutgeb:2021mpu}, {the $a=3$ sector is unmodified}, but the $a=(0,8)$ equations of motion are changed to
\bea\label{psiAnHW}
    \partial_z\left(\frac1z \partial_z \psi^a_{A,n}(z)\right)+\frac1z m_{A,n}^2 \psi^a_{A,n}(z) \nonumber
    \\
-\frac{g_5^2( M_{ab}^2+\delta_{0a} \delta_{0b}\tilde{Y}_0^2)}{z^3} \psi^b_{A,n}(z)=0.
\eea
In each case the axial vector TFF is given by\footnote{Note that in the flavor-symmetric case considered in \cite{Leutgeb:2019gbz}
$A$ without flavor index was defined differently,
corresponding to (the then $a$-independent) $A^a$ here.
}
\bea
\label{eq:axtff}
    &&A_n(Q_1^2, Q_2^2)=\text{tr}(t^a \mathcal{Q}^2) A_n^{a}(Q_1^2, Q_2^2),
\eea
with
\begin{equation}\label{TFFAn} 
    A_n^a(Q_1^2,Q_2^2) = \frac{2g_5}{Q_1^2} \int_0^{z_0} \!\!\! dz \left[ \frac{d}{dz} \mathcal{J}(Q_1,z) \right]
    \mathcal{J}(Q_2,z) \psi^{A,a}_n(z).
\end{equation}
At large $Q^2$ we obtain as in \cite{Leutgeb:2019gbz,Leutgeb:2021mpu}
\be
    A_n^a(Q^2 (1+w), Q^2 (1-w)) \to \frac{g_5^2 F^a_{A,n}}{Q^4} f_A(w) ,
\ee
with the decay constants 
\be
F_n^a=\left.-g_5^{-2}\partial_z\psi^{a}_{A,n}/z\right|_{z\to0}
\ee
and the asymmetry function
\begin{equation}
    f_A(w) = \frac1{w^4} \left[ w(3-2w)  + \frac12 (w+3)(1-w)\ln\frac{1-w}{1+w}\right],
\end{equation}
in agreement with the asymptotic form derived from QCD in \cite{Hoferichter:2020lap}.

At $Q_1^2=Q_2^2=0$,
the axial vector TFF in \eqref{eq:axtff} is related to the form factor $ F^{(1)}_{\mathcal{A}_n\gamma^*\gamma^*}(0,0)$ defined in \cite{Pascalutsa:2012pr} via
$m_{A_n}^{-2}  F^{(1)}_{\mathcal{A}_n\gamma^*\gamma^*}(0,0) = \frac{N_c}{4\pi^2} A_n(0,0).$ (See the Appendix of \cite{Leutgeb:2019gbz} for more details.)

The most general expression for axial vector amplitudes has actually one further asymmetric structure function \cite{Pascalutsa:2012pr,Roig:2019reh,Zanke:2021wiq}, which is set to zero in the holographic model and whose phenomenological importance has not yet been established; see Ref.~\cite{Zanke:2021wiq} for
a compilation of the available phenomenological information.

\section{Results}

\subsection{Parameter settings}

As one of the input data which we fit, we take the $\rho$ meson mass\footnote{A shortcoming of the minimal HW models considered here is that the strange quark mass modifies the vector meson masses too little compared to reality: $\rho$, $\omega$ and $\phi$ mesons are degenerate, the mass of $K^*$ is raised to only 0.79 GeV.}
$m_\rho=\gamma_{0,1}z_0^{-1}=2.40483\ldots z_0^{-1}$, where $\gamma_{0,1}$
is the first zero of the Bessel function $J_0$. Following Ref.~\cite{Abidin:2009aj}, we have chosen $z_0^{-1}=0.3225$ GeV corresponding to $m_\rho=775.556$ MeV. This fixes the location
of the hard wall, $z_0$, and $\Lambda$ in
the expression for $\alpha_s$.
The coupling $g_5$ is either set by the leading-order OPE result \eqref{g5LO} or the slightly reduced value \eqref{g5Frho} obtained by fitting the $\rho$ meson decay constant, where the TFFs reach only 89.4\% of the OPE and Brodsky-Lepage limits,
thereby coming closer to next-to-leading order results at moderately large, experimentally relevant energy scales.

The isospin-symmetric quark mass parameter $m_q$ and the chiral condensate parameter $\sigma$ are chosen such that $m_\pi=134.97$ MeV and $f_\pi=92.21$ MeV
\cite{FlavourLatticeAveragingGroupFLAG:2021npn}; the strange quark mass
parameter $m_s$ is chosen such that \cite{Brunner:2015oga}
\bea
m_K^2&=&\frac12(m_{K_\pm}^2+m_{K_0}^2)-
\frac12(m_{\pi_\pm}^2-m_{\pi_0}^2)\nonumber\\
&&= (495.007\mathrm{MeV})^2
\eea
in order to minimize isospin-breaking contributions.

For the two choices of $g_5$, we consider the model with and without a gluon condensate parameter $\Xi_0$. When $\Xi_0=0$ (referred to as model version v0 in the following), we obtain predictions for $m_\eta$
and $m_\eta'$ that are around 10\% lower than the real-world values, in accordance
with Ref.~\cite{Katz:2007tf} who had omitted to turn on a nonzero $\Xi_0$. 
Fitting $\Xi_0$ such that $(1-m^\mathrm{th}_{\eta}/m^\mathrm{exp}_{\eta})^2+(1-m^\mathrm{th}_{\eta'}/m^\mathrm{exp}_{\eta'})^2$ is minimized (model v1), $m_\eta$
and $m_\eta'$ can be matched at the percent level, as shown in
Table \ref{tab:etas}. 
In the four versions of our model, we have chosen a large value of  $\kappa=700$, in order to be in the regime where the dependence on $\kappa$
is rather weak.

\subsection{Decay constants and photon coupling}

Up to the slightly different choice of $f_\pi$, the results for the mesons in the isotriplet sector, where $\Xi_0$ does not play a role, are identical to the
HW1m model presented in \cite{Leutgeb:2021mpu} for $g_5=2\pi$. Table \ref{tab:pia1}
generalizes this to the case where $g_5$ is fitted to match $F_\rho$.

In Tables \ref{tab:etas} and \ref{tab:AV}, detailed results for the two versions v0 and v1 are given for the first few pseudoscalar and axial vector modes in the isosinglet sector, showing their mixing behavior in the decay constants $f^8$, $f^0$, $f_G$ for the $\eta$'s, and $F_A^8$, $F_A^0$ for the $f_1$'s, as well
as in the coupling to real photons given by $F(0,0)$ and $A(0,0)$, respectively.
All results are given in units of GeV raised to the appropriate power; note that
the mass dimension of $f^8$ and $f^0$ is 1, but that of $f_G$ is 3; $F(0,0)$ and
$A(0,0)$ have mass dimensions $-1$ and $-2$, respectively. 

Mixing is in fact energy dependent in the holographic model because
the components of the holographic wave functions depend nontrivially
on the holographic coordinate $z$ which corresponds to inverse energy.
The mixing angles read from decay constants thus differ from those read
from the components of the photon coupling. Moreover, the pseudoscalar
mixings are different when determined from $\eta$ or $\eta'$
(and similarly in the case of $f_1$ and $f_1'$). Indeed, a phenomenological need for
an energy-dependent mixing in the case of $\eta$ and $\eta'$
has been argued for in \cite{Escribano:1999nh,Escribano:2005qq}.

The pseudoscalars $\eta$, $\eta'$ and a third ground-state $\eta''$ meson arise from mixing of flavor octet and flavor singlet degrees of freedom with the pseudoscalar glueball $G$, each followed by an infinite tower of excited states.
The ground state modes are dominantly flavor octet, singlet, and
glueball judging from the corresponding decay constants evaluated at $z\to0$,
while the first excited triplet $\eta^{(3)}$ to $\eta^{(5)}$ shows a more
involved mixing behavior. 
The decay constants for $\eta$ and $\eta'$ agree reasonably well with the
recent lattice results of Ref.~\cite{Bali:2021qem}, where also pseudoscalar matrix
elements have been evaluated. Our results for $f_G$ correspond to $\sqrt{N_f/2}a$ in \cite{Bali:2021qem} and also agree reasonably well. The ratio $a_{\eta'}/a_\eta$ is between 2 and 2.5 depending on the renormalization scale. This is better in line with our model v1 that includes a nonzero gluon condensate, where $f_{G,\eta'}/f_{G,\eta}=2.60$ and 2.66 for the two choices of $g_5$, while
model v0 has 1.46 and 1.30.

Without gluon condensate (v0), the results for $F(0,0)$ show rather poor agreement
with experimental results for the $\eta$ meson with deviations of around 30\%, while
those for $F_{\eta'\to\gamma\gamma}(0,0)$ are much better. With gluon condensate (model v1), where the masses of $\eta$ and $\eta'$ agree with experimental data
at the percent level,
both couplings turn out to agree remarkably well with the experimental values.

For isosinglet axial vector mesons (Table \ref{tab:AV}), both model versions predict
generally too high values of $f_1$ and $f_1'$ masses (+8\% to +28\% compared to PDG data \cite{ParticleDataGroup:2022pth}). The $f_1$ and $f_1'$ mesons are obtained as
dominantly flavor octet and flavor singlet, respectively.
In the holographic model, the mixing angle is an energy or $z$ dependent
quantity. In the case of the $f_1$ mesons, it is usually
extracted from equivalent photon decay rates at zero virtuality, where the experimental results from
the L3 experiment read \cite{Achard:2001uu,Achard:2007hm}
\be
\tilde\Gamma_{\gamma\gamma}=\left\{
\begin{array}{cc}
    3.5(8) \;\text{keV} &  \text{for}\;f_1=f_1(1285)\\
    3.2(9) \;\text{keV} & \text{for}\;f_1'=f_1(1420)
\end{array}
\right.
.
\ee
With the definition
\be
f_1=\cos\theta_A f^0+\sin\theta_A f^8
\ee
and the assumption that
$\tilde\Gamma_{\gamma\gamma}\propto m_A$, one has
\be
\tan^2(\theta_A-\arcsin\frac13)=\frac{m_{f_1}\tilde\Gamma_{\gamma\gamma}^{f_1'}}{m_{f_1'}\tilde\Gamma_{\gamma\gamma}^{f_1}},
\ee
leading to  \cite{Hoferichter:2020lap}
$\theta_A=62(5)^\circ$, superficially agreeing with model version v0.
However, in the holographic model we have 
\be
\tilde\Gamma_{n,\gamma\gamma}
= \frac{\pi\alpha^2 m_A}{12} \left[ \frac{N_c m_A^2}{4\pi^2} A_n(0,0) 
\right]^2 
\sim m_A(m_A/\Lambda)^4,
\ee
resulting in $\theta_A=56(5)^\circ$ 
for the experimental value, which
does not fit to the results for either v0 or v1, the latter
disagreeing even more than the former.\footnote{It would be interesting to revisit this issue in other holographic
QCD models, in particular ones that are closer to a string-theoretic top-down construction such as the models of Ref.~\cite{Casero:2007ae,Arean:2013tja,Giannuzzi:2021euy}.}
While the mixing angle depends rather strongly on $\Xi_0$,
the combination $\sqrt{[A^8(0,0)]^2+[A^0(0,0)]^2}$ changes only slightly 
between models v0 and v1, and it
is also close to the value of $A(0,0)$ in the isotriplet sector,
as well as to the same quantity in the chiral hard-wall model
\cite{Leutgeb:2019gbz}, $(21.04\,\text{GeV})^{-2}$.
Matching $A(0,0)$ with $\tilde\Gamma_{\gamma\gamma}\propto m_A(m_A/\Lambda)^4$
to the L3 results
leads to a value of 15.2(2.0) GeV$^{-2}$ so that the holographic results, which
read 20-21 GeV$^{-2}$ when $g_5=2\pi$ and 19-20 GeV$^{-2}$ for the reduced $g_5$, are somewhat too high for $f_1$ and $f_1'$, but
not excluded for $a_1$, for which Ref.~\cite{Roig:2019reh} has a concordant
estimate of 19.3(5.0) GeV$^{-2}$.

\subsection{Transition form factors}

For the HLBL contribution of single mesons to $a_\upmu$, their singly and doubly virtual TFFs are of
critical importance. 

\begin{figure}
\bigskip
\centerline{$Q^2 F_{\pi^0\gamma^*\gamma}(Q^2,0)$ [GeV]\hfill}
\includegraphics[width=0.42\textwidth]{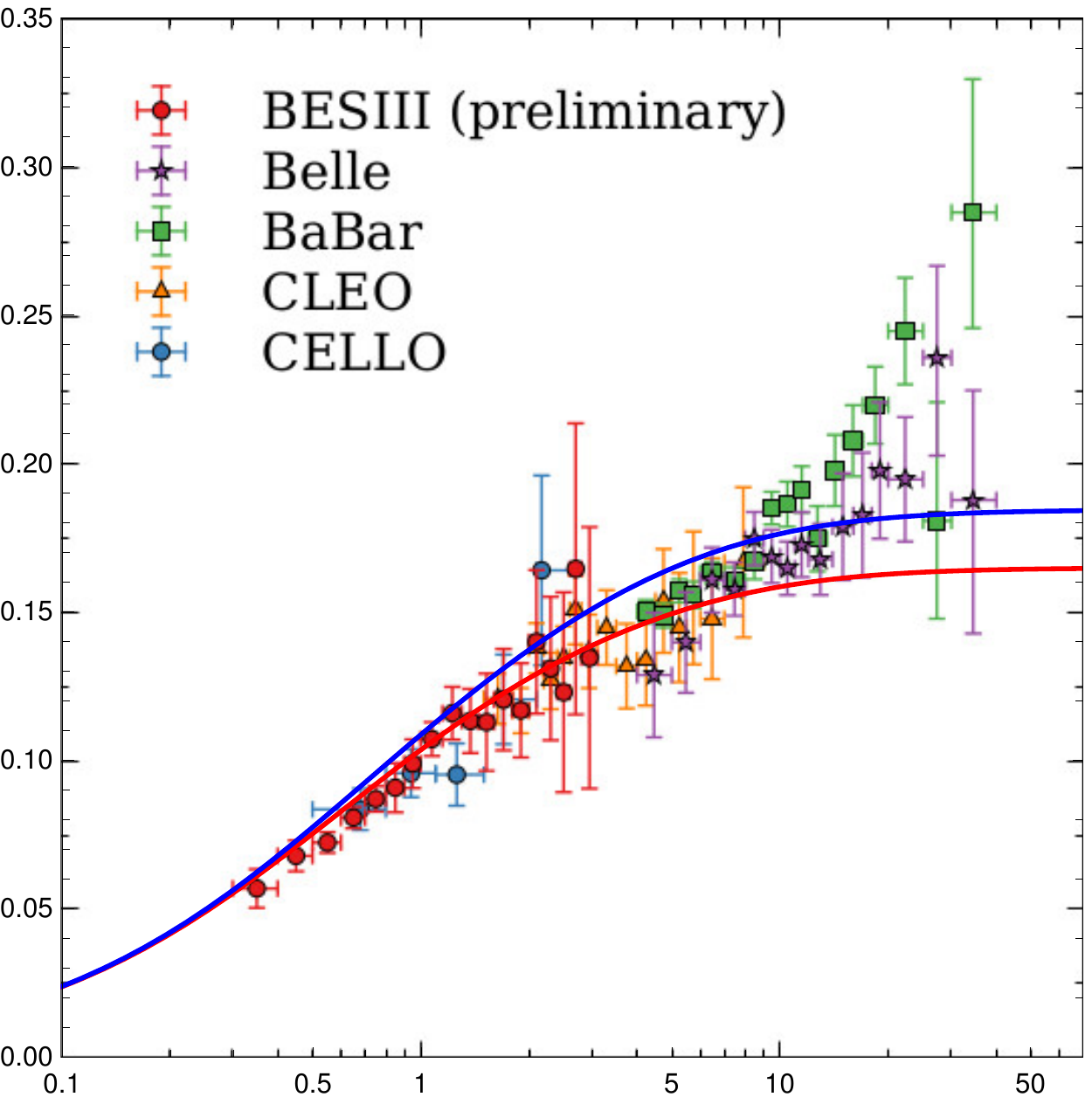}
\centerline{$Q^2$ [GeV$^2$]}

\caption{Holographic results for the single virtual TFF $Q^2 F(Q^2,0)$ for $\pi^0$, plotted on top of experimental data as compiled in Fig.~53 of Ref.~\cite{Aoyama:2020ynm}
for $g_5=2\pi$ (OPE fit, blue) and the reduced value (red)
corresponding to a fit of $F_\rho$.
 (For $\pi^0$ results for the model with and without gluon condensate coincide.)}
\label{fig53comp}
\end{figure}

As in the chiral HW model \cite{Leutgeb:2019zpq},  we find excellent agreement
of the singly virtual result for the pion TFF with available experimental data, see Fig.~\ref{fig53comp}. At virtualities relevant for $a_\upmu$, the
results with $g_5$ fitted to $F_\rho$, where the asymptotic limit
is 89.4\% of the Brodsky-Lepage value, seem to give the best match.

\begin{figure}
\bigskip
\includegraphics[width=0.42\textwidth]{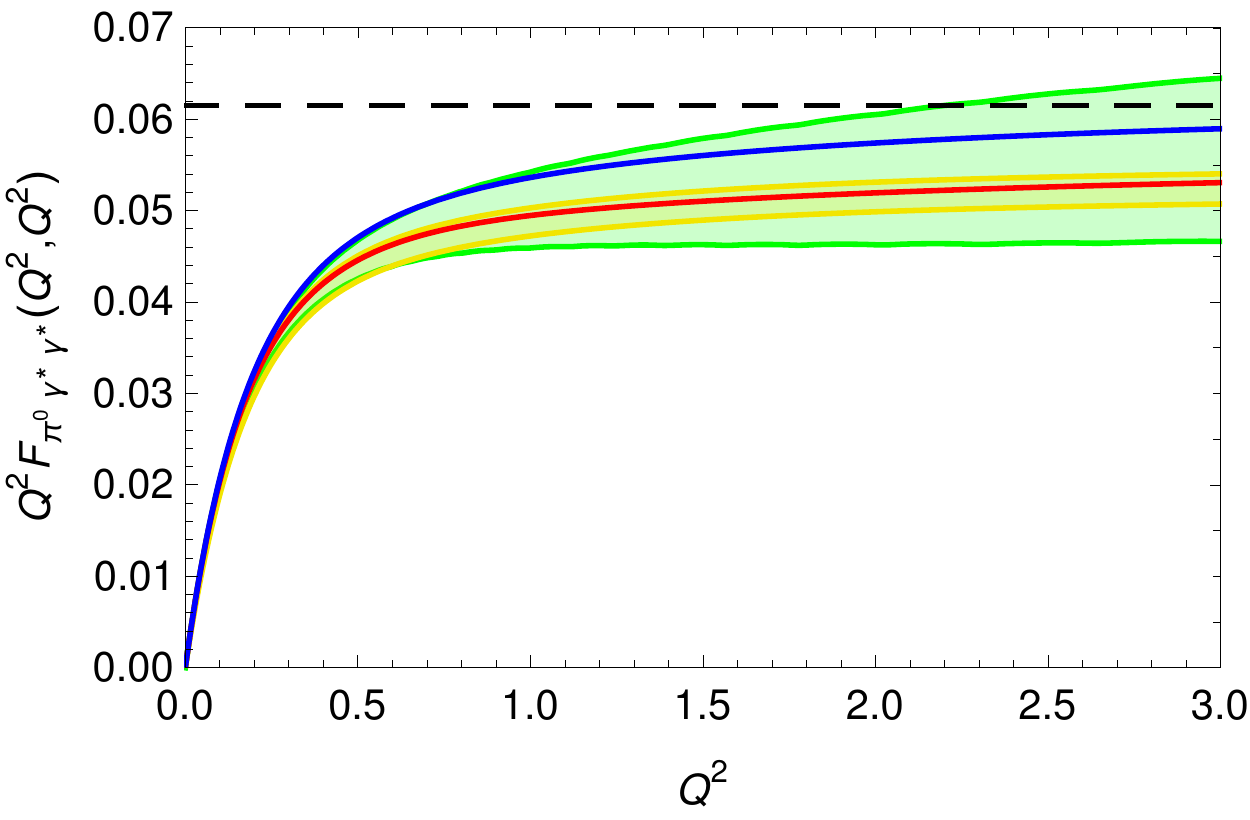}
\caption{Holographic results for the doubly virtual $F_{\pi^0\gamma^*\gamma^*}$ compared to the dispersive result of Ref.~\cite{Hoferichter:2018kwz} (green band) and the lattice result of Ref.~\cite{Gerardin:2019vio} (yellow band); the OPE limit given by the dashed horizontal line. The upper full line (blue) corresponds to $g_5=2\pi$ (OPE fit), the lower (red) one to the reduced value $g_5$ ($F_\rho$-fit). (Here the two versions with and without gluon condensate coincide.)
}
\label{figpi0TFFd}
\end{figure}

For the symmetric doubly virtual TFF the comparison is made with
the dispersive result of Ref.~\cite{Hoferichter:2018kwz} and the lattice result of Ref.~\cite{Gerardin:2019vio} 
in Fig.~\ref{figpi0TFFd}. Both choices of $g_5$ are within the error band
of the dispersive result, while the result for the reduced $g_5$ is
also within the error band of the lattice result and moreover 
happens to coincide with
the central values of the dispersive approach within line thickness of the plot
throughout the entire range of $Q^2$.

\begin{figure}
\bigskip
\centerline{$Q^2 F_{P\gamma^*\gamma}(Q^2,0)$ [GeV]\hfill}
\includegraphics[width=0.42\textwidth]{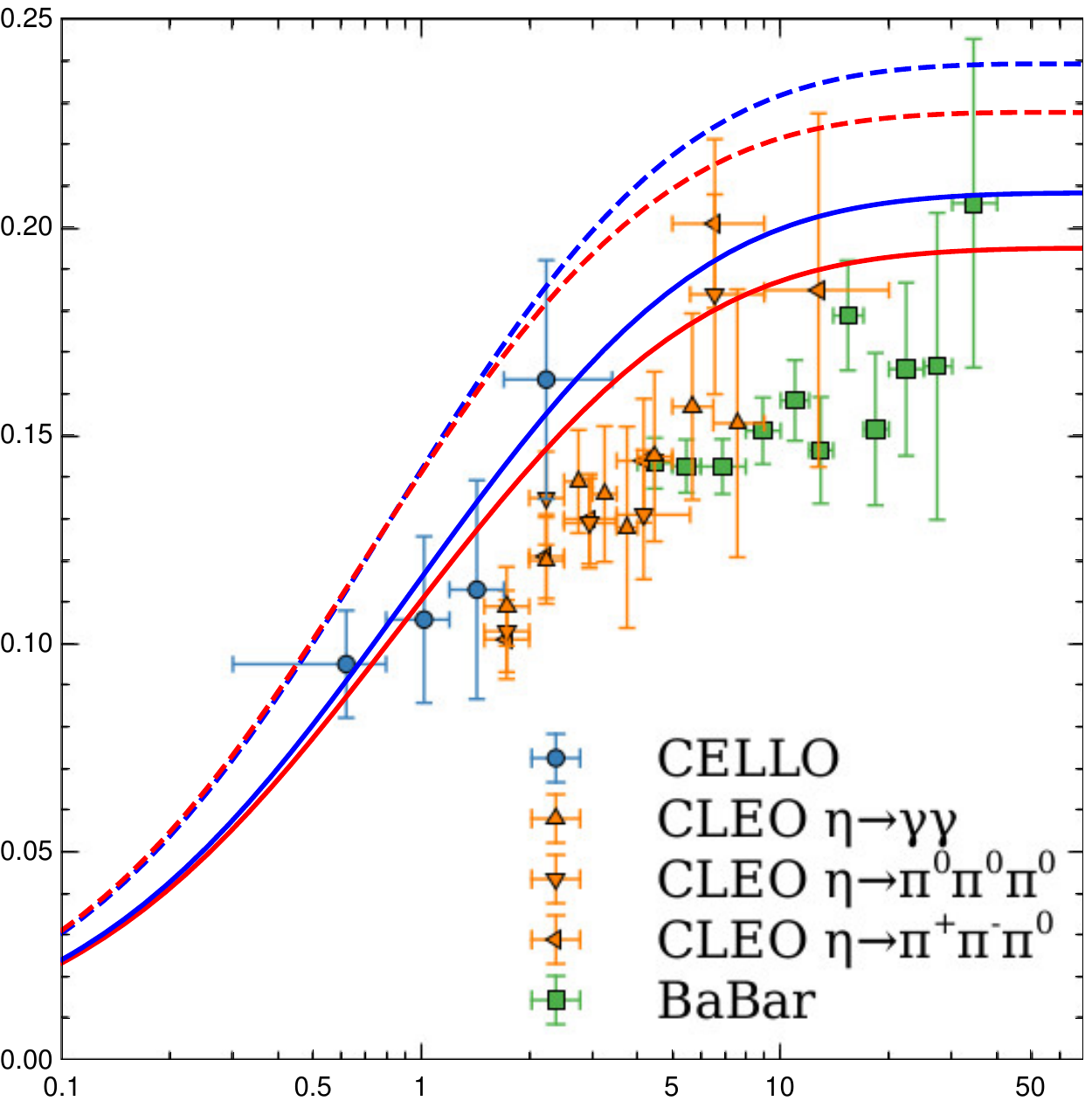}
\includegraphics[width=0.42\textwidth]{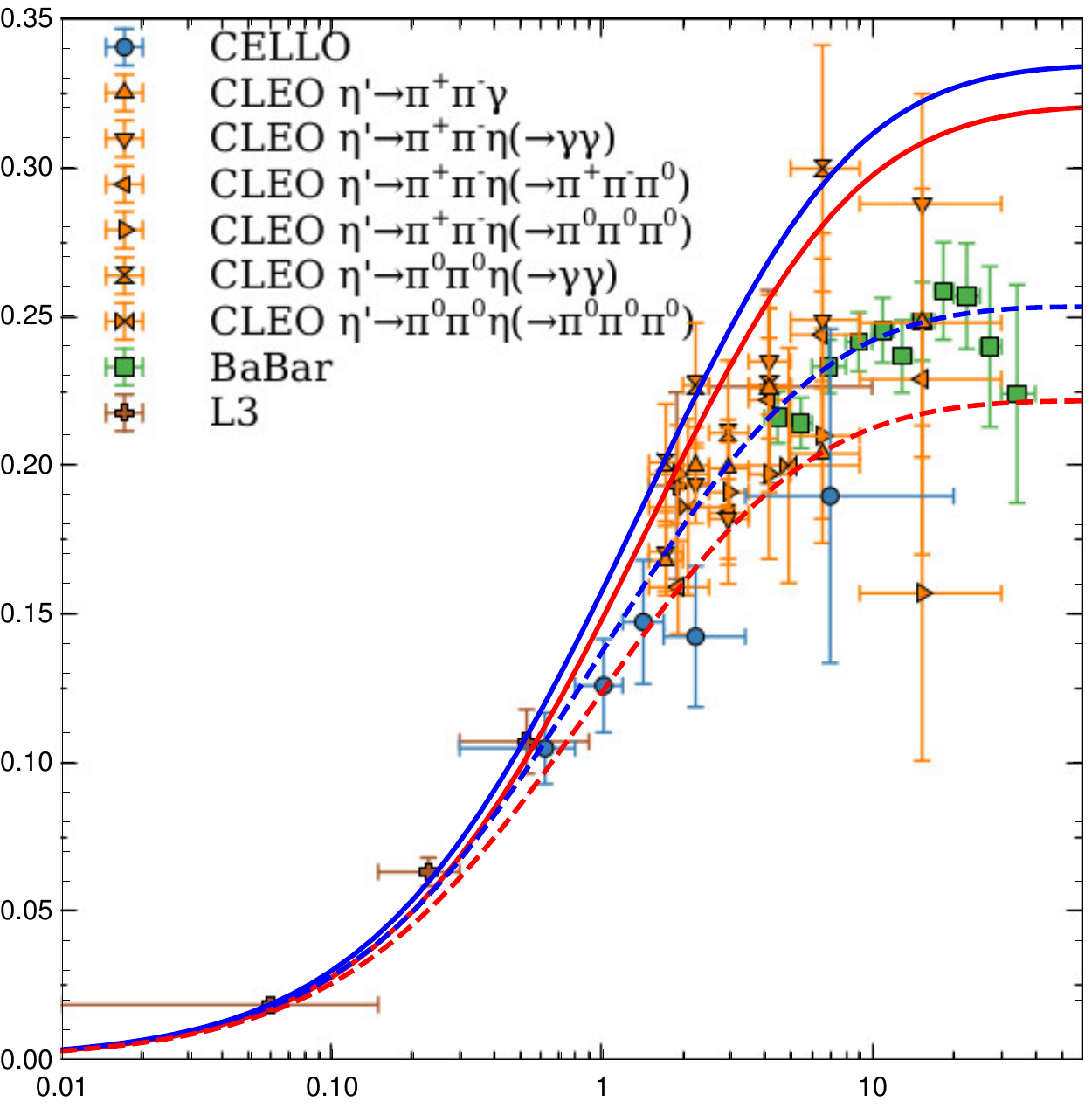}
\centerline{$Q^2$ [GeV$^2$]}

\caption{Holographic results for the single virtual TFF $Q^2 F(Q^2,0)$ for $\eta$ and $\eta'$ plotted on top of experimental data as compiled in Fig.~54 of Ref.~\cite{Aoyama:2020ynm}
for $g_5=2\pi$ (OPE fit, blue) and the reduced value (red)
corresponding to a fit of $F_\rho$. Full lines are with gluon condensate (version v1), dashed lines without (v0).
}
\label{fig54comp}
\end{figure}

With $\eta$ and $\eta'$ mesons, there is a rather strong
dependence on the parameter $\Xi_0$ representing a gluon condensate.
With this parameter turned on, the masses of $\eta$ and $\eta'$ can be
matched to percent level accuracy, and the resulting prediction for
$F_{P\gamma\gamma}(0,0)$ is then in complete agreement with experiment for $g_5=2\pi$ (see Table \ref{tab:etas}), while with reduced $g_5$ this value is
slightly underestimated in the case of $\eta'$.
For the singly virtual TFF of $\eta$, only the results with nonzero $\Xi_0$ are
close to the experimental data, see Fig.~\ref{fig54comp}. They match those at low $Q^2$ quite well,
but are generally larger at higher virtualities.
In the case of $\eta'$, all model versions agree with the low-$Q^2$ date
due to L3, while at higher $Q^2$ the results without gluon condensate
agree with more of the data points, but only with unreduced $g_5=2\pi$.

\begin{figure}
\bigskip
\includegraphics[width=0.42\textwidth]{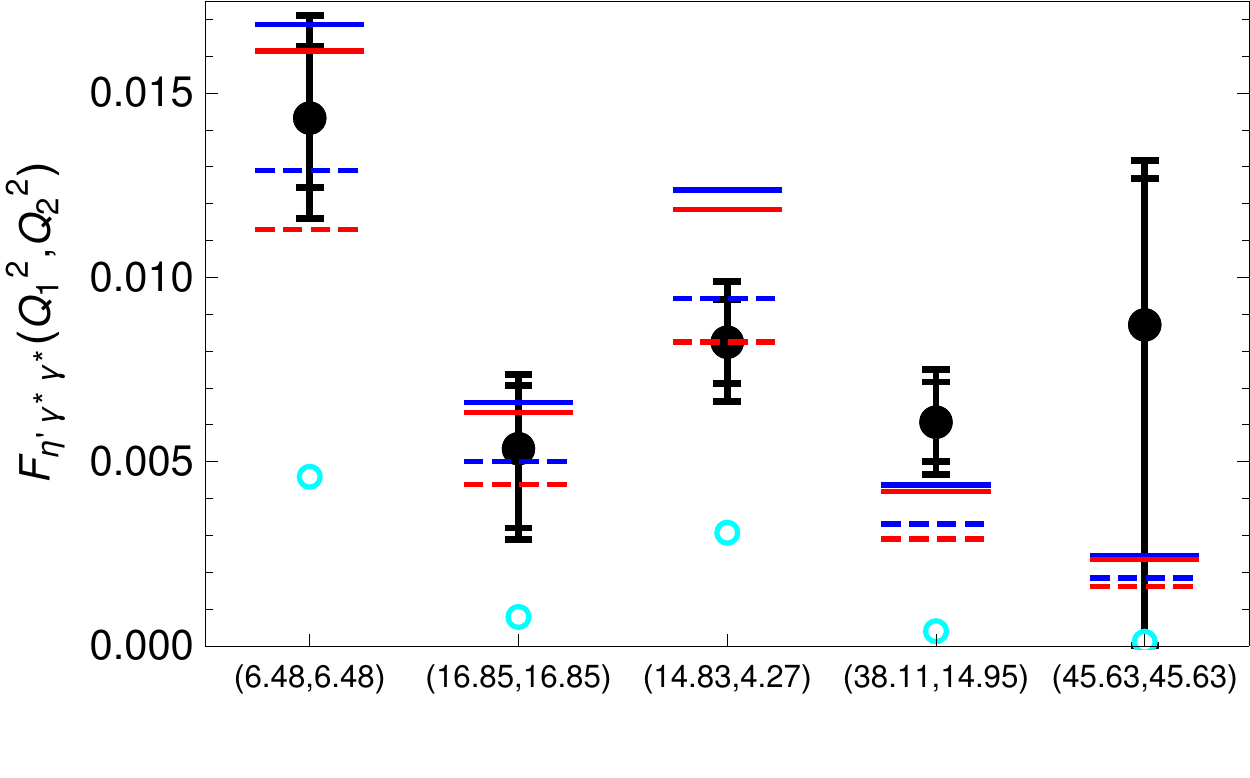}
\caption{Holographic results for the doubly virtual $F_{\eta'\gamma^*\gamma^*}$ compared to BABAR data points (black) and a simple VMD model fitted with singly virtual data (cyan circles) \cite{BaBar:2018zpn}. Full lines are with gluon condensate (version v1), dashed lines without (v0); blue color corresponds to $g_5=2\pi$ (OPE fit) and red to the reduced value $g_5$ ($F_\rho$-fit).
}
\label{figBBetacomp}
\end{figure}

In contrast to the case of $\pi^0$, there are also several experimental data points for the doubly virtual TFF of $\eta'$. As opposed to the simple VMD model 
considered in \cite{BaBar:2018zpn} and represented by the cyan circles in Fig.~\ref{figBBetacomp}, the holographic results are within 1-2 standard deviations.
For the lowest virtualities $Q^2_1=Q_2^2=6.48 \text{GeV}^2$, which are the most significant for $a_\upmu$, all versions of the model come close to the experimental result. With gluon condensate, the agreement is better with the reduced $g_5$, whereas without gluon condensate, a reduction of $g_5$ to fit $F_\rho$ moves the prediction slightly outside the error bar.

All in all, the model with gluon condensate and reduced $g_5$ seems to
be the optimal choice regarding pseudoscalar TFFs.

\begin{table*}[] 
    \centering
\begin{tabular}{lcccccccp{12pt}ccccccc}
\toprule
& \multicolumn{6}{c}{$g_5^2=(2\pi)^2$}                                    & & \multicolumn{6}{c}{$g_5^2=0.894(2\pi)^2$}    \\ 
              & $\pi^0$ &  $\pi^*$ & $a_1$ & $a_1^*$ & {$a_1^{**}$} & $a_1^{***}$ & $a_1^{****}$  & &  $\pi^0$ &  $\pi^*$ & $a_1$ & $a_1^*$ & {$a_1^{**}$} & $a_1^{***}$ & $a_1^{****}$  \\ 
              \hline
$m$ & 0.135$^*$ & 1.891 & 1.363 & 2.137 & {2.987} & 3.935 & 4.916 && 0.135$^*$ & 1.841  & 1.278 & 2.047 & {2.936} & 3.902 & 4.891 \\
$f\;\vee\;F_A/m_A$ & 0.09221$^*$ & 0.00157 & 0.175 & 0.204 & {0.263} &0.311 & 0.330 && 0.09221$^*$ &0.00173&0.173 &0.217&0.280 & 0.329 & 0.330\\
$F(0,0)\vee {A^3(0,0)}$  &0.277 &-0.203 & 20.96& 3.31& {-0.336}& 2.16 & 0.370 && 0.276&-0.199&19.46&4.87&-0.413 & 2.05 & 0.325 \\ \hline
$a_\upmu\times 10^{11}$ & 66.1 & 0.73 & 7.83 & 1.24 & {0.44} & 0.28 & 0.11 &&
63.4 & 0.71 & 7.09 & 1.47 & 0.42 & 0.26 & 0.10 \\
\botrule
\end{tabular}
    \caption{Results for pseudoscalar and axial vector mesons in the isotriplet sector (the gluon condensate parameter $\Xi_0$ does not play a role here). All quantities in units of (powers of) GeV.}\label{tab:pia1}
\end{table*}

\begin{table*} 
\centering\bigskip
\begin{tabular}{lccccccp{12pt}cccccc}
\toprule
\textbf{(v0)}         & \multicolumn{6}{c}{$\Xi_0=0$}                                 && \multicolumn{6}{c}{$\Xi_0=0$}                                 \\ 
& \multicolumn{6}{c}{$g_5^2=(2\pi)^2$}                                    & & \multicolumn{6}{c}{$g_5^2=0.894(2\pi)^2$}    \\
              & $\eta$  & $\eta'$ & $G/\eta''$     & $\eta^{(3)}$ & {$\eta^{(4)}$}  & $\eta^{(5)}$  && $\eta$  & $\eta'$ & $G/\eta''$      &  $\eta^{(3)}$ & $\eta^{(4)}$ & $\eta^{(5)}$ \\ \colrule
$m$  & 0.513     & 0.840     & 1.862 & 1.999    & {2.257}  & 2.705  && 0.503     & 0.819   & 1.764  & 1.948     & {2.207}  & 2.638  \\
${m}-{m^\mathrm{exp}}$ & -6.4\% & -12.3\% & & & & && -8.2\% & -14.5\% & & \\
$f^8$         & 0.0917 & -0.0565  & 0.00197 & 0.0266 & {0.0121} &0.0080 && 0.0902 & -0.0624 & 0.00405 & 0.0293  & {0.0132}&0.00837 \\
$f^0$         & 0.0394 & 0.0945 & -0.0212 & -0.00823 & {-0.0390} &0.0362 && 0.0446 & 0.0952 & -0.0224 & -0.00802  & {-0.0416} &0.0337 \\
$f_G$         & -0.0264  & -0.0385  & 0.0674 & -0.0400  & {0.154} &-0.310 && -0.0265 & -0.0344  & 0.0600 & -0.0454   & {0.156} &-0.280 \\
$F^8(0,0)$         & 1.46   & -0.674  & 0.177 & -1.18    & {0.00233} &0.236 && 1.41   & -0.737 & 0.0640  & -1.16     & {0.0239}&0.241 \\
$F^0(0,0)$         & 0.776  & 1.42  & 0.169 & 0.0383 & {1.08} &0.229 && 0.828  & 1.34   & 0.00310  & 0.00492 & {1.10}&0.253   \\
$F(0,0)$           & 0.351  & 0.322 & 0.0629 & -0.103   & {0.293} &0.0851 && 0.361  & 0.295  & 0.00700 & -0.110    & {0.302}&0.0922  \\
${F}-{F^\mathrm{exp}}$ & +28(2)\% & -6(2)\% & & & & && +32(2)\% & -14(2)\% & & \\
\hline
$a_\upmu\times 10^{11}$ & 32.8 & 15.7 & 0.055 & 0.14 & {0.79} & 0.16 && 34.0 & 13.3 & 0.003 & 0.16 & 0.85 & 0.16 \\
\botrule
\end{tabular}

\medskip
\begin{tabular}{lccccccp{12pt}cccccc}
\toprule
\textbf{(v1)}         & \multicolumn{6}{c}{$\Xi_0=0.01051$}                                 && \multicolumn{6}{c}{$\Xi_0=0.01416$}                                 \\ 
& \multicolumn{6}{c}{$g_5^2=(2\pi)^2$}                                    & & \multicolumn{6}{c}{$g_5^2=0.894(2\pi)^2$}    \\
              & $\eta$  & $\eta'$ & $G/\eta''$     & $\eta^{(3)}$ & {$\eta^{(4)}$} & $\eta^{(5)}$  && $\eta$  & $\eta'$ & $G/\eta''$      &  $\eta^{(3)}$ & $\eta^{(4)}$ & $\eta^{(5)}$ \\ \colrule
$m$ & 0.557     & 0.950     & 1.992    & {2.390}    & 2.954 & 3.214 && 0.561     & 0.947     & 1.943    & 2.428  & 2.914 & 3.317 \\
${m}-{m^\mathrm{exp}}$  & +1.7\% & -0.8\% & & {} & & && +2.4\% & -1.1\% & & \\
$f^8$         & 0.101  & -0.0385  & -0.0267  & {0.0116} & -0.0228 &-0.0049 && 0.103  & -0.0393  & -0.0299 & 0.0112 & -0.0253&-0.00767 \\
$f^0$         & 0.0272 & 0.113  & 0.0049  & {-0.0492}  &-0.00115 &-0.0214 && 0.0298 & 0.121  & 0.00761  & -0.0522 & 0.00320&-0.0128 \\
$f_G$         & -0.0298   & -0.0774  & 0.053  & {0.233}  & 0.1483 &0.269 && -0.0313  & -0.0821  & 0.048 & 0.260 & 0.1236& 0.214 \\
$F^8(0,0)$         & 1.55   & -0.431   & 1.19   & {-0.0478}  &-0.887 & 0.167&& 1.53   & -0.442   & 1.149  & -0.0312 & -0.877&0.129 \\
$F^0(0,0)$         & 0.468  & 1.40   & 0.0051 & {0.904}  & 0.0300 &0.0867 && 0.444  & 1.31   & -0.000026  & 0.837 & 0.0307&0.130 \\
$F(0,0)$           & 0.276  & 0.340  & 0.116  & {0.241}  & -0.0772 &0.0397 && 0.268  & 0.313  & 0.111  & 0.225 & -0.0760& 0.0477 \\
${F}-{F^\mathrm{exp}}$ & +1(2)\% & -0(2)\% & & {} & & && +2(2)\% & -8(2)\% & & \\
\hline
$a_\upmu\times 10^{11}$ & 19.3 & 16.9 & 0.19 & {0.53} & 0.043 & 0.008 && 17.6 & 14.9 & 0.18 & 0.45 & 0.039 & 0.007 \\
\botrule
\end{tabular}
\caption{Results for the isoscalar pseudoscalar sector, for the model with (v1) and without (v0) gluon condensate, and for two choices of $g_5$: $g_5=2\pi$ corresponding to matching the vector correlator to the LO UV-behavior in QCD, and the reduced value corresponding to a fit of $F_\rho$. All dimensionful quantities in units of (powers of) GeV.}
\label{tab:etas}
\end{table*}

\begin{table}[] 
    \centering
\begin{tabular}{lccp{12pt}cc}
\toprule
\textbf{(v0)}         & \multicolumn{2}{c}{$\Xi_0=0$}                                 && \multicolumn{2}{c}{$\Xi_0=0$}                                 \\ 
& \multicolumn{2}{c}{$g_5^2=(2\pi)^2$}                                    && \multicolumn{2}{c}{$g_5^2=0.894(2\pi)^2$}    \\
               & $f_1$ &  {$f_1'$}  && $f_1$ &  {$f_1'$}      \\ 
              \colrule
$m$ & 1.460 & {1.651} && 1.388 & {1.598} \\
${m}-{m^\mathrm{exp}}$ & +14\% & +16\% && +8\% & +12\% \\
$F_A^8/m$  & 0.163 & -0.0732 && 0.165 & -0.0627 \\
$F_A^0/m$  & 0.0743 & 0.169 && 0.0690 & 0.180 \\
$A^8(0,0)$  & 19.27 & -8.649 && 18.38 & -7.194 \\
$A^0(0,0)$  & 8.676 & 19.21 && 7.310 & 18.62 \\
$\theta_A$  & $65.8^\circ$ & -24.2$^\circ$ && 68.3$^\circ$ & -21.1$^\circ$ \\
$A(0,0)$  & 4.22 & 4.40 && 3.76 & 4.37 \\
$m^*$ & 2.241 & 2.614 && 2.147 & 2.561 \\
$m^{**}$ & 3.056 & 3.580 && 2.999 & 3.535 \\ \hline
$a_\upmu\times 10^{11}$ & 11.0 & {10.8} && 9.08 & 11.0 \\
$a^*_\upmu\times 10^{11}$ & 0.61 & 1.50 && 0.62 & 1.54 \\
$a^{**}_\upmu\times 10^{11}$ & 0.18 & 1.08 && 0.16 & 0.99 \\
$a^{***}_\upmu\times 10^{11}$ & 0.09 & 0.42 && 0.08 & 0.39 \\
$a^{****}_\upmu\times 10^{11}$ & 0.04 & 0.27 && 0.03 & 0.25 \\
\botrule
\end{tabular}

\medskip
\begin{tabular}{lccp{12pt}cc}
\toprule
\textbf{(v1)}         & \multicolumn{2}{c}{$\Xi_0=0.01051$}                                 && \multicolumn{2}{c}{$\Xi_0=0.01416$}                                 \\ 
& \multicolumn{2}{c}{$g_5^2=(2\pi)^2$}                                    && \multicolumn{2}{c}{$g_5^2=0.894(2\pi)^2$}    \\
               & $f_1$ &  {$f_1'$}  && $f_1$ &  {$f_1'$}      \\ 
\colrule
$m$ & 1.481 & 1.810  &&  1.410 & 1.820 \\
$m-m^\mathrm{exp}$ & +15\% & +27\% && +10\% & +28\% \\
$F_A^8/m_A$  &0.176 &-0.0299 &&0.176 & -0.0167\\
$F_A^0/m_A$  & 0.0365&0.201 && 0.0292&0.219 \\
$A^8(0,0)$  &20.77 &-3.842 && 19.58&-2.556 \\
$A^0(0,0)$  &3.857 &20.07 && 2.690& 19.00\\
$\theta_A$  &79.5$^\circ$ &-10.8$^\circ$ &&82.2$^\circ$ &-7.7$^\circ$ \\
$A(0,0)$  &3.05 & 5.09&& 2.62&4.93 \\
$m^*$ & 2.246 & 2.862 && 2.153 & 2.891 \\
$m^{**}$ & 3.058 & 3.869 && 3.004 & 3.907 \\ \hline
$a_\upmu\times 10^{11}$ & 5.71 & 14.3 && 4.34 & 13.6 \\
$a^*_\upmu\times 10^{11}$ & 0.36 & 1.01 && 0.33 & 0.91 \\
$a^{**}_\upmu\times 10^{11}$ & 0.11 & 1.11 && 0.05 & 0.99 \\
$a^{***}_\upmu\times 10^{11}$ & 0.01 & 0.33 && 0.02 & 0.24 \\
$a^{****}_\upmu\times 10^{11}$ & 0.01 & 0.28 && 0.05 & 0.15 \\
\botrule
\end{tabular}
    \caption{Results for the isoscalar axial vector sector, for the model with (v1) and without (v0) gluon condensate, and the two choices $g_5$(OPE fit) and
    $g_5$($F_\rho$-fit). Here $\theta_A\equiv\arctan(A^8(0,0)/A^0(0,0))$
    for both $f_1$ and $f_1'$, {and $A(0,0)=\text{tr}(t^a\mathcal{Q}^2)A^a(0,0)=[A^8(0,0)+\sqrt{8}A^0(0,0)]/6\sqrt{3}$}. All dimensionful quantities are given in units of (powers of) GeV. In the $a_\upmu$ contributions, about 58\% are due to the longitudinal part of the axial vector meson propagator, which contributes to the MV constraint.}
    \label{tab:AV}
\end{table}

\subsection{HLBL contribution to $a_\upmu$}

Tables \ref{tab:pia1} and \ref{tab:etas} include also the individual contributions
of the listed pseudoscalar and axial vector meson modes to $a_\upmu$, which are
collected in Table \ref{tab:total} for the model with nonzero gluon condensate (v1) with $g_5=2\pi$ (OPE-fit)
and the reduced value \eqref{g5Frho} from fitting the $\rho$ meson decay. Only with
the extra parameter $\Xi_0$ for the gluon condensate, the predictions for $F_{P\gamma\gamma}(0,0)$ and masses of $\eta$ and $\eta'$ match experimental data with good accuracy. With reduced $g_5$ ($F_\rho$-fit), the predictions for
$a_\upmu^{\pi^0}$ and $a_\upmu^{\eta'}$ are
extremely close to the central values adopted by the White Paper
\cite{Aoyama:2020ynm}, and those for $\eta$ agree within $1\sigma$.

The holographic model also includes a third ground-state $\eta$ meson, which we called $\eta''$, the result of mixing with the pseudoscalar glueball $G$.
It contributes only $0.2\times 10^{-11}$, but
there is also a whole tower of excited $\eta$ modes, which together with excited pion modes contribute around $1.5\times 10^{-11}$ so that the total pseudoscalar poles prediction
for model v1($F_\rho$-fit) is close to the upper end of the WP prediction, whereas
the result for model v1(OPE fit) is 2.5$\sigma$ higher.

The main aim of this study is of course the experimentally less well constrained axial vector meson contribution, which in holographic QCD has been shown to
take into account the Melnikov-Vainshtein short-distance constraint \cite{Leutgeb:2019gbz,Cappiello:2019hwh}, also away from the chiral limit \cite{Leutgeb:2021mpu}. The holographic result thus presents an alternative
estimate of the combined contribution of axial vector mesons, for which
the WP estimate is $6(6)\times 10^{-11}$, and of short-distance contributions\footnote{In the symmetric high-energy limit, the holographic results
for the HLBL scattering amplitude have the correct dependence on $Q^2$, but reproduce the OPE value only at the level of 81\% when $g_5=2\pi$, where the asymmetric
MV limit is saturated fully \cite{Cappiello:2019hwh,Leutgeb:2021mpu}.},
estimated in the WP as $15(10)\times 10^{-11}$. With errors added linearly,
the WP value is at $21(16)\times 10^{-11}$.

It is difficult to estimate errors for any holographic result, but we expect
our results for $a_\upmu$ to be in good shape despite some deviations in its
ingredients.
The holographic results for axial vector mesons have turned out to overestimate the masses of $f_1$ and $f_1'$ by 8-28\%, where the models with gluon condensate have the higher deviations. On the other hand, all our models have an equivalent real photon coupling $A(0,0)$ 
that is 20-28\% too large compared to the value derived from L3 data for $f_1$ and $f_1'$, albeit
in good agreement with the estimate of Ref.~\cite{Roig:2019reh} for $a_1(1260)$.
The mixing angles for $f_1$ and $f_1'$ are poorly predicted, and even worse
when the gluon condensate is turned on. However, the prediction for the amplitude $\sqrt{(A^8)^2+(A^0)^2}$
appears to be fairly robust and only weakly dependent on $\Xi_0$. 
A different modeling of the gluon condensate could perhaps lead to better
predictions for the mixing with similar overall amplitude.
Our summary in Table \ref{tab:total} therefore lists the presumably more reliable
combined contribution of $f_1$ and $f_1'$.
Since the contribution to $a_\upmu$ decreases with increasing axial vector meson mass by approximately two inverse powers while the amplitude $A$ enters quadratically, we expect that the errors in the predictions of both will largely cancel out, so that the holographic results can still be a reasonably good
prediction for the axial vector meson contributions to $a_\upmu$.
For our favored model v1($F_\rho$-fit), the contribution from the
ground-state axial vector mesons is $a_\upmu^{a_1+f_1+f_1'}=25.0\times 10^{-11}$,
about 4 times the WP estimate. The contribution from $f_1+f_1'$ is 2.5 times that
of $a_1$, somewhat reduced from the flavor-U(3)-symmetric value of 3 that was
assumed in our previous estimates in Ref.~\cite{Leutgeb:2021mpu}. For this contribution, Pauk and Vanderhaeghen \cite{Pauk:2014rta}
have estimated a value of only $6.4(2.0)\times 10^{-11}$, much smaller than
our holographic prediction of $17.9\times 10^{-11}$. Besides the differences in $A(0,0)$ and the mass parameters, a crucial difference of the TFF assumed in \cite{Pauk:2014rta} is that it is obtained from a factorized ansatz
that unlike the holographic result does not have the correct asymptotic
behavior \cite{Hoferichter:2020lap} in the doubly virtual case, where it
falls off as $1/Q^{4}$ instead of $1/Q^2$.

In the holographic models, the excited axial vector mesons ensure agreement
with the longitudinal (Melnikov-Vainshtein) short-distance constraint. This
constraint derived from the axial anomaly is satisfied to 100\% in the model v1(OPE fit), and to 89.4\% in the case of v1($F_\rho$-fit). The latter should
provide a better approximation at large but still physically relevant energy
scales, where typically $\sim 10\%$ of next-to-leading order pQCD corrections apply \cite{Melic:2002ij,Bijnens:2021jqo}. 

In the chiral HW1 model and in the U(3)-symmetric massive HW1m model that we have investigated in Refs.~\cite{Leutgeb:2019gbz,Leutgeb:2021mpu}, we have obtained 9.2 and $9.4\times10^{-11}$
from excited axial vectors, where 25\% are due to $a_1$ by U(3) symmetry.
The contribution of excited $a_1$'s in our present models are essentially
the same as in the HW1m model (up to a slightly different fit value of $f_\pi)$,
but the excited isoscalars remain below the extra factor of 3 expected from U(3) symmetry.\footnote{In order to approximate the sum of contributions from the infinite tower of axial vector mesons we have used the observation that in the chiral HW models as well as in the HW1m model the infinite series of contributions can be roughly approximated by a geometric one with $a_{n+1}/a_n\approx 0.6$ for $n>2$. The full sum can thus be approximated by multiplying the last contribution of a truncated sum by a factor of 1/(1-0.6)=2.5. In the case of excited pseudoscalars, which do not contribute to the longitudinal short-distance constraint \cite{Leutgeb:2021mpu}, the contributions drop much more quickly. Our results for those are obtained simply from the sum of the first few modes.} Instead, the latter provide only 1.6 and 1.4 times the contributions from excited $a_1$'s in the case of v1(OPE fit) and v1($F_\rho$-fit), respectively. 

The total contribution from axial vector
mesons is thus significantly smaller than the estimates we have come up with
in the flavor-symmetric case of Ref.~\cite{Leutgeb:2021mpu}: 33.7 and 30.5$\times 10^{-11}$ for the two choices of $g_5$ (instead of 40.8 and 38.8$\times10^{-11}$ for HW1m and HW1m with reduced $g_5$, respectively). Comparing this to the combined estimate of axial vector mesons and short-distance contributions in the WP, $21(16)\times10^{-11}$, we find values that are about 50\% higher, but well within the estimated error.

\begin{table}[t]  
\bigskip
\centering
\begin{tabular}{lccc}
\toprule
$a_\upmu^{...}\times 10^{11}$ & v1(OPE fit) &  v1($F_\rho$-fit) & WP \\
 \hline
 $\pi^0$ & 66.1 &  63.4 & 62.6$^{+3.0}_{-2.5}$ \\
 $\eta$ & 19.3 &  17.6 & 16.3(1.4) \\
 $\eta'$ & 16.9 &  14.9 & 14.5(1.9) \\
 $G/\eta''$ & 0.2 &  0.2 \\
 $\sum_{PS^*}$ & 1.6 & 1.4 \\[4pt]
 \hline
 PS poles total & 104 &  97.5 & 93.8(4.0) \\
 \hline
 $a_1$ & 7.8 &  7.1 \\
 $f_1+f_1'$  & 20.0 &  17.9 \\
 $\sum_{a_1^*}$ & 2.2 &  2.4 \\
 $\sum_{f_1^{(')*}} $ & 3.6 &  3.0 \\[4pt] 
\hline
 AV+LSDC total & 33.7 &  30.5 & 21(16) \\
 \hline
 total & 138 & 128 & 115(16.5) \\
 \botrule
\end{tabular}
    \caption{Summary of the results for the different contributions to $a_\upmu$ in comparison with the White Paper \cite{Aoyama:2020ynm} values.}
    \label{tab:total}
\end{table}

\section{Conclusion}

In this paper, we have upgraded our previous studies of the HLBL contribution
in HW AdS/QCD models to 2+1 flavors with strange quark mass $m_s>m_u=m_d$
plus a Witten-Veneziano mass for the flavor singlet degree of freedom
generated by interaction terms involving a pseudoscalar glueball with
the latter that implement the anomalous Ward identities of the $U(1)_A$ symmetry
in the line of Ref.~\cite{Katz:2007tf,Schafer:2007qy}.

In holographic QCD, the Melnikov-Vainshtein constraint on the HLBL scattering
amplitude is naturally satisfied, to the same degree that TFFs satisfy the
Brodsky-Lepage and OPE limits. All these are saturated at the level of 100\%
for the standard value of $g_5=2\pi$ in HW1 models.\footnote{The simpler Hirn-Sanz (HW2) model, which omits the bifundamental scalar $X$, reaches 62\% when $f_\pi$ and $m_\rho$ are fitted.} However, because these
models do not involve a running coupling in the UV, the UV-limits of TFFs
are approached too quickly, likely
leading to overestimated HLBL contributions to $a_\upmu$.
Next-to-leading-order gluonic corrections in pQCD suggest a reduction
by about 10\% at large but still experimentally relevant virtualities.
Precisely such a correction is obtained by fitting $g_5$ such that the
decay constant of the $\rho$ meson is matched instead of the OPE result
for the vector correlator. In Ref.~\cite{Leutgeb:2022cvg}, we have found
that this also moves the $N_f=2$ result of HW AdS/QCD models for the HVP contribution much closer to the dispersive results \cite{Davier:2019can,Keshavarzi:2019abf}.

In Ref.~\cite{Leutgeb:2019gbz,Leutgeb:2021mpu} we have shown that the MV short-distance constraint
is realized by the infinite tower of axial vector mesons, with the excited
axial vector mesons adding about a third of the contribution from the ground-state
axial vectors in the flavor-symmetric case. A much smaller contribution
comes from excited pseudoscalars, which do not contribute to the
longitudinal short-distance behavior at leading order.

In our present study with $U(1)_A$ anomaly included, where we have
obtained a remarkably accurate fit of the masses of $\eta$ and $\eta'$ mesons
as well as of their $F_{P\gamma\gamma}(0,0)$ values when including a
nonzero gluon condensate that was omitted in \cite{Katz:2007tf},
we have found a reduction of the ratio 3:1 for the isoscalar:isotriplet
contributions of axial vector mesons to about 2.5:1. For excited
mesons (axial vector as well as pseudoscalar), we have obtained an even
more pronounced reduction, which reduces our prediction for the $a_\upmu$
contribution of axial vector mesons in the U(3)-symmetric case
from around 41 and 39$\times 10^{-11}$ to 33.7 and 30.5$\times 10^{-11}$
for $g_5$(OPE) and $g_5$($F_\rho$-fit), respectively.
These values are above the estimate of the White Paper \cite{Aoyama:2020ynm}
for the contribution of (ground-state) axial vector mesons plus
short-distance constraints, but still within the error given there.
The pseudoscalar contributions obtained in our model v1($F_\rho$-fit)
agree completely with the WP results for $\pi^0$, $\eta$, and $\eta'$,
however this model also has a contribution of $1.6\times 10^{-11}$ from excited pseudoscalars, where the tower of $\eta$'s mixes with a pseudoscalar glueball.
The complete contribution from summing pseudoscalar and axial vector contributions
is approximately $128\times10^{-11}$, which we consider our currently best estimate obtained from AdS/QCD; it
thus turns out to be close to (but below) the upper end of the corresponding WP estimate.

\begin{acknowledgments}
We would like to thank
Gilberto Colangelo, Martin Hoferichter, Bastian Kubis, Elias Kiritsis, and Pablo Sanchez-Puertas for helpful discussions.
J.~L.\ and J.~M.\ have been supported by the Austrian Science Fund FWF, project no. P33655,
and by the FWF doctoral program
Particles \& Interactions, project no. W1252-N27.
\end{acknowledgments}

\raggedright
\bibliographystyle{JHEP}
\bibliography{hlbl}

\end{document}